\documentclass[journal]{IEEEtran}
\usepackage[T1]{fontenc}
\usepackage[latin9]{inputenc}
\usepackage{amsthm}
\usepackage{amsmath}
\usepackage{graphicx}
\usepackage{color}
\usepackage{cite}
\usepackage{algpseudocode}
\usepackage{algorithm}
\usepackage{amssymb}
\usepackage{subfigure}
\usepackage{esint}
\usepackage{algpseudocode}
\usepackage{epstopdf}
\usepackage{multirow}
\usepackage{makecell}
\usepackage{float}
\usepackage{lipsum}
\usepackage{balance}
\usepackage{tikz}
\usetikzlibrary{shapes ,arrows, positioning} 
\usepackage{adjustbox}
\usepackage{bbm}
\usepackage{hyperref}
\usepackage{bm}

\newtheorem*{definition*}{Definition}
\newtheorem{remark}{Remark}

\theoremstyle{plain}
\theoremstyle{plain}

\newcommand{\comment}[1]{}

\IEEEoverridecommandlockouts
\usepackage{colortbl}
\usepackage{hhline}
\begin{document}
%
\title{Learning to Reconfigure: Using Device Status to Select the Right Constrained Coding Scheme}
\author{
\IEEEauthorblockN{Do\u{g}ukan \"{O}zbayrak and Ahmed Hareedy,~\IEEEmembership{Member,~IEEE}}\vspace{-1.7em}
\thanks{Do\u{g}ukan \"{O}zbayrak and Ahmed Hareedy are with the Department of Electrical and Electronics Engineering, Middle East Technical University, 06800 Ankara, Turkey (e-mail: dogukan.ozbayrak@metu.edu.tr and ahareedy@metu.edu.tr).
This work was supported in part by the T\"{U}B\.{I}TAK 2232-B International Fellowship for Early Stage Researchers.}
}
\markboth{}%
{}
%
\IEEEtitleabstractindextext{%
\begin{abstract}
In the age of data revolution, a modern storage~or transmission system typically requires different levels of protection. For example, the coding technique used to fortify data in a modern storage system when the device is fresh cannot be the same as that used when the device ages. Therefore, providing reconfigurable coding schemes and devising an effective way to perform this reconfiguration are key to extending the device lifetime. We focus on constrained coding schemes for the emerging two-dimensional magnetic recording (TDMR) technology. Recently, we have designed efficient lexicographically-ordered constrained (LOCO) coding schemes for various stages of the TDMR device lifetime, focusing on the elimination of isolation patterns, and demonstrated remarkable gains by using them. LOCO codes are naturally reconfigurable, and we exploit this feature in our work. Reconfiguration based on predetermined time stamps, which is what the industry adopts, neglects the actual device status. Instead, we propose offline and online learning methods to perform this task based on the device status. In offline learning, training data is assumed to be available throughout the time span of interest, while in online learning, we only use training data at specific time intervals to make consequential decisions. We fit the training data to polynomial equations that give the bit error rate in terms of TD density, then design an optimization problem in order to reach the optimal reconfiguration decisions to switch from a coding scheme to another. The objective is to maximize the storage capacity and/or minimize the decoding complexity. The problem reduces to a linear programming problem. We show that our solution is the global optimal based on problem characteristics, and we offer various experimental results that demonstrate the effectiveness of our approach in TDMR systems. This approach can also be extended to other data storage or transmission systems.

\end{abstract}
}
\maketitle
\IEEEdisplaynontitleabstractindextext
%
\IEEEpeerreviewmaketitle
\vspace{-0.5em}
\section{Introduction}\label{sec_intro}

The demand for reliability, density, and adaptivity in modern data storage and transmission systems is increasing as time goes by. Due to the ultra high data volumes used in these systems as well as their strict reliability requirements, effects of device aging become more severe. Therefore, fortification requirements for stored or transmitted data increase throughout the device lifetime. For example, a coding scheme for data protection to be used in the early lifetime of a storage device may not be effective late in the device lifetime, and performance degradation might occur consequently, which necessitates a stronger coding scheme for that regime \cite{ahh_otloco}. Thus, designing different coding schemes for different lifetime stages of the device and reconfiguring to switch between these schemes have become essential for ensuring consistently reliable performance and/or extending device lifetime. Performing reconfiguration between the coding schemes optimally is crucial because uninformed reconfiguration can either unjustifiably reduce storage capacity (transmission rate) or fail to provide sufficient data fortification. For example, consider a data storage device where the objective is to maximize storage capacity under error performance constraints. If reconfiguration here occurs earlier than the optimal point, the system might operate with unnecessary redundancy and waste capacity. Conversely, if it occurs later than the optimal point, performance degradation is likely as constraints are violated. Therefore, with optimal reconfiguration between coding schemes, the system achieves the best trade-off between reliability and storage capacity throughout the device lifetime.

Machine learning methods have found a broad range of applications within areas such as coding theory and modern data storage. In particular, these methods can be used to facilitate improving the performance of coding schemes and in storage systems. In \cite{ml1}, how advances in storage technologies and in machine learning support each other is discussed. For example, machine learning techniques can benefit from new storage technologies to handle large amounts of data and large models efficiently. In \cite{ml2}, deep learning is used to improve the error-correction decoder performance by proposing a cost function that leads to increasing the focus of training on incorrectly decoded bits by adjusting the gradients. A learning-based approach for modeling Flash memory channels where data is subjected to constrained coding is introduced in \cite{ml3}. Zheng and Siegel accurately replicate pattern-dependent error statistics to accelerate training and improve accuracy in reconstructing read voltages in order to support the design of efficient error-correction coding schemes for Flash memories. In \cite{ml4}, Mei et al. introduce a deep transfer learning approach that leverages knowledge from a device wear state to quickly fine-tune a neural network detector for future conditions. This method significantly cuts down the amount of training data needed. Furthermore, there are ongoing research and valuable contributions in terms of learning techniques to aid the code construction and/or the decoder design. In \cite{ml5}, a framework that uses artificial intelligence for constructing error-correction codes is presented. This framework employs reinforcement learning and genetic algorithms as code builders within a general constructor-evaluator framework. In \cite{ml6}, Sandell and Ismail propose using neural networks trained on offline data across various operating conditions to estimate log-likelihood ratios for the decoding of low-density parity-check (LDPC) codes efficiently, avoiding complex real-time computations. In \cite{ml7}, Xiao et al. frame the design of finite-alphabet iterative decoders for LDPC codes as a training problem for recurrent quantized neural networks. In \cite{ml8}, Qin et al. show how convolutional neural networks can be used to improve data-detection performance for two-dimensional magnetic recording. Other relevant works include \cite{ml9} and \cite{ml10}.

While our approach can be extended to other coding techniques and modern systems, we focus here on reconfiguring between constrained coding schemes designed for the emerging two-dimensional magnetic recording (TDMR) technology \cite{wood_tdmr, chan_tdmr}. The TDMR technology promises storage densities of up to $10$ terabits per square inch \cite{victora_10tb} and throughputs of more than $10$ gigabits per second \cite{vasic_hist}. One of the main advantages of TDMR is that it relies on advanced data processing methods to increase storage capacity without requiring new magnetic materials \cite{shayan_tdmr}. As the TDMR device ages, the TD density unintentionally increases (as we will explain later), causing the interference effects to increase among adjacent tracks. This behavior motivates the need for reconfiguring the coding schemes to maintain performance requirements.

Constrained codes are channel codes that are used to prevent error-prone data patterns from being written to storage media or being transmitted in transmission standards \cite{marcus_book}. Constrained codes have been used in a wide range of applications to protect data for various technologies since Shannon discussed constrained systems in 1948 \cite{shan_const}. They are employed in one-dimensional magnetic recording (ODMR) devices to control transition separation \cite{ahh_loco} as well as in data-transmission computer standards \cite{sridhara_ctalk} to improve reliability. In Flash memory devices, constrained codes mitigate inter-cell interference and can extend the device lifetime \cite{veeresh_mlc}. TD constrained coding has various applications as well \cite{sharov_TCon}. In 1970, Tang and Bahl presented the first design of run-length-limited (RLL) constrained codes based on lexicographic indexing \cite{tang_bahl}. This result was later followed by other results on the design of RLL codes based on lexicographic indexing \cite{gu_lex}. Adler, Coppersmith, and Hassner introduced a finite-state machines-based (FSM-based) systematic design method for constrained codes in 1983 \cite{ach_fsm}, a method that was subsequently adopted by many researchers \cite{siegel_mr}. In 2022, we introduced a general design framework for lexicographically-ordered constrained (LOCO) codes \cite{ahh_general}, which is inspired by Cover's 1973 work on enumerative coding \cite{cover_lex}. Using this method, we developed multiple LOCO codes for various applications \cite{ahh_rr} and demonstrated that their reconfigurability can be exploited to transition between coding schemes as error-prone patterns change with device aging, ensuring consistent reliability levels over time.

To forbid the error-prone TD patterns in TDMR, we recently introduced rate-wise optimal LOCO codes in \cite{ahh_general} and \cite{ahh_otloco} that improve the performance of the TDMR device adopting a wide read head. The wide read head accesses three adjacent down tracks simultaneously, which enables one-dimensional non-binary constrained coding solutions. These optimal LOCO codes incur relatively high encoding-decoding complexity and are relatively more susceptible to codeword-to-message error propagation due to their high alphabet size and message lengths. To overcome the aforementioned drawbacks, we designed simple LOCO coding schemes that are defined over alphabets of smaller sizes compared with the optimal LOCO codes \cite{ahh_simpleloco}. These simple LOCO coding schemes offer lower encoding-decoding complexity, reduced latency, and minimal error propagation, with a cost of modest rate penalty. The complexity of any LOCO coding scheme is dictated by the size of the adder performing the encoding-decoding rule \cite{ahh_general}. As they are encoded and decoded via an adder regardless from the constraint, LOCO codes are reconfigurable, i.e., the code can be changed by altering the inputs to the adder, which are the code cardinalities \cite{ahh_loco, ahh_general}. 

The idea of reconfiguration between coding schemes can be potentially used in different applications and for different classes of coding techniques; from data storage to wireless communications, from Flash memories to magnetic recording, and from constrained codes to error-correction codes. We are interested in designing an optimal reconfiguration approach to switch between LOCO codes for TDMR in this work. In the TDMR device, interference effects increase as the device ages. A direct consequence of this behavior is that error-prone patterns change as the device ages, which motivates the need for reconfiguring the used LOCO coding scheme. The main goal we want to achieve is to perform the reconfiguration optimally, that is, to select switching points between the coding schemes in a way that maximizes the storage capacity, extends the device lifetime, and/or minimizes the encoding-decoding complexity. In this work, we propose offline and online learning methods to determine which coding schemes should be used as well as the reconfiguration points effectively. In offline learning, the complete dataset covering the entire device lifetime of interest is assumed to be available, whereas training data collected at specific separated time intervals is utilized to make sequential reconfiguration decisions in online learning. We then fit the training data to polynomial equations that give the bit error rate (BER) in terms of the TD density to be used in the optimization problems we will design.

In this paper, our contributions are:
\begin{enumerate}
\item Offline learning: We present three different optimization problems, or setups, depending on the application needs. The first aims to maximize storage capacity, the second aims to find the best compromise between maximizing storage capacity and minimizing complexity, and the third aims to maximize storage capacity while keeping the average adder size under a predefined threshold. Using curve fitting, we derive closed-form error rate functions, which are used to specify constraints ensuring reliable performance. We prove that these problems are \textit{convex} and they reduce to \textit{linear programming} problems. We present solutions for these optimization problems and show the results, which reveal interesting findings and gains over straightforward methods.
\item Online learning: We present four online learning setups that differ with respect to the associated training intervals, their lengths and their time/density locations. The objective is that the TDMR device controller anticipates the optimal switching points (given the available data) in the respective near future, through solving our optimization problems, to maximize storage capacity during the device operation. We design and solve the relevant optimization problems to engineer effective online learning procedures. We show the results obtained for these aforementioned setups.
\item Simulation result: We compare offline learning and the online learning methods. We also compare the methods developed in this paper to a straightforward idea that divides the device lifetime equally between different coding schemes, and to a reconfiguration idea developed in a previous work \cite{ahh_otloco}. These comparisons demonstrate the notable gains achieved by our methods.
\end{enumerate}

The rest of the paper is organized as follows. In Section~\ref{sec_motiv}, we motivate the need to find optimal ways of switching between LOCO codes and present the practical TDMR system model we use. In Section~\ref{offline}, we introduce our offline learning method with its different optimization problems, derive the solutions, and analyze the findings. In Section~\ref{online}, we present our online learning method with its different setups, derive the solutions, and discuss the results. In section~\ref{sim_res}, we compare the results obtained via offline learning, online learning, the aforementioned straightforward idea, and the idea used in~\cite{ahh_otloco}. We also show performance plots to compare different setups. In Section~\ref{sec_conc}, we conclude the paper and state some future~work.
%
%
%
%
\section{TDMR System Model and Motivation}\label{sec_motiv}

In this section, we first motivate the need for reconfiguration between different LOCO coding schemes throughout the TDMR device lifetime, and then briefly discuss the TDMR system model we adopt to perform our simulations.

In Section~\ref{sec_intro}, we stated that error-prone data patterns in the TDMR system change as the device ages. This behavior is confirmed via the simulations we performed over a practical TDMR model. In particular, using the uncoded setting demonstrates that the error profile dynamics change as the TD density of the device increases, i.e., as the device gets older \cite{ahh_otloco}. The simulations we performed demonstrate that plus isolation (PIS) patterns are the dominant error-prone patterns when the device is fresh. However, as the device ages, new error-prone patterns emerge and surpass PIS patterns in the error-profile share, and these are called incomplete plus isolation (IPIS) patterns. PIS and IPIS patterns are $3 \times 3$ patterns on the TD grid of TDMR tracks, which we will explain shortly. The fact that the error profile changes motivates the need to reconfigure the constrained coding scheme to prevent detrimental patterns and provide reliable performance. We define PIS and IPIS patterns collectively as rotated T isolation (RTIS) patterns. The reconfigurability property of LOCO codes allows us to use constrained coding schemes preventing PIS patterns, namely optimal plus LOCO (OP-LOCO) and simple plus LOCO (SP-LOCO) coding schemes, at the early and intermediate stages of the TDMR device lifetime, then to switch to constrained coding schemes preventing RTIS patterns, namely optimal T LOCO (OT-LOCO) and simple T LOCO (ST-LOCO) coding schemes, as the device gets older. We skip the technical details regarding the construction of these codes, which can be found in \cite{ahh_otloco} and \cite{ahh_simpleloco}, to focus on the novel results. We will introduce the code parameters (rates and adder sizes) as well as the BER performance plots of interest, which are given in Fig~\ref{fig_perf1}.

Determining the reconfiguration points between coding schemes optimally is pivotal to fully utilize the device storage capacity while extending its lifetime. To effectively switch between coding schemes, we design different reconfiguration approaches and consider different objective functions and constraints. In particular, we offer offline and online learning approaches based on data fitting and optimization. Moreover, our objective functions and constraints, collectively the optimality measures, are the storage capacity or/and the average encoding-decoding complexity in addition to the device lifetime. Observe that our online learning method gives the optimal solution conditioned on the available training data. Such a solution is, in general, sub-optimal if the data is available throughout the device lifetime.

Now, we discuss the practical TDMR model we use to perform our simulations and extract the data. Denote the TD channel (read-head) impulse response duration at half the amplitude in the cross-track and the down-track directions as $PW_{50,\textup{CT}}$ and $PW_{50,\textup{DT}}$, respectively. Also denote the track width and bit period by $TW$ and $BP$, respectively. The sweep metric used to generate performance plots is the \textit{TD channel density}, $D_{\textup{TD}} = (PW_{50,\textup{CT}} \times PW_{50,\textup{DT}})/(TW \times BP)$. This metric is also what we use to represent the device lifetime stage or the device age. The relation between the TD density and the device lifetime is explained in Remark~\ref{rmk_density_time} explicitly. The model adopts level-based signaling, with two levels having the same magnitude and opposite polarities. In the TDMR setup, we adopt a wide-read head, separating down tracks into disjoint groups of $3$ tracks each. Interference in the cross-track direction from a group into another group is negligible \cite{chan_tdmr}, \cite{bd_tdmr}. Thus, the TD channel impulse response is discretized as a $3 \times 3$ matrix representing the intersection of $3$ adjacent down tracks in the same group with $3$ consecutive cross tracks.

\begin{remark} \label{rmk_density_time}
As the device ages, the TD channel impulse response gets wider spatially. When the TD impulse response gets wider spatially, the widths at half the amplitude get bigger. Given that the underlying technology remains unchanged, it is reasonable to conclude that the TD density $D_{\textup{TD}}$ increases as the device ages. Therefore, it is important to note that throughout the paper, time (age) and density can be used interchangeably. TD density here is used as a measure of time in a way similar to using program/erase cycles as a measure of time in Flash memory systems.
\end{remark}

Interference on the same down track (inter-symbol interference or ISI) and on the same cross track (inter-track interference or ITI) can result in the level at the central position of any $3 \times 3$ TD grid changing its sign, which results in an error upon reading, if this level is \textit{isolated}. Isolation here means the level at the central position is surrounded on the TD grid by $4$ levels at Manhattan distance $1$ or by $3$ levels at Manhattan distance $1$ with the complementary sign, and the sets of equivalent $3 \times 3$ binary patterns resulting in such isolation are the set of PIS patterns and the set of IPIS patterns, respectively. PIS patterns and IPIS patterns collectively form the set of RTIS patterns. More details about the error-prone patterns can be found in \cite{ahh_otloco}.

\textbf{Writing setup:} We generate random binary input messages. Then, we encode each message into the corresponding codeword. Each symbol in the codeword along with the free data (if any) are converted into a $3 \times 1$ column of binary bits according to the mapping-demapping of the related coding scheme \cite{ahh_otloco}, \cite{ahh_simpleloco}. Before writing, level-based signaling is applied. The schemes we are using in this work are OP-LOCO, SP-LOCO, OT-LOCO, and ST-LOCO coding schemes. OP-LOCO and OT-LOCO codes are rate-wise optimal LOCO codes, whereas SP-LOCO and ST-LOCO coding schemes are simple LOCO coding schemes as we discussed earlier. OP-LOCO and SP-LOCO coding schemes are designed to forbid PIS patterns, while OT-LOCO and ST-LOCO coding schemes are designed to forbid RTIS patterns.

\textbf{Channel setup:} Our baseline channel model is the TDMR model in \cite{mohsen_tdmr}, which is a Voronoi model. Here, we only consider media noise/interference. We modify this model such that it is suitable for a wide read head. We do not scale $TW$ and $BP$ with the square root of the code rate here.

\textbf{Reading setup:} Outputs of the channel are read based on a hard decision applied to the value of each grid entry. We provide BER plots in this work. Bit errors, as the name tells, are just counted when the hard-decision value of the channel output does not match the input value to the channel ($+1$ or $-1$). Observe that the additional protection on the upper and lower tracks (due to the wide read head) makes the average amount of interference required to cause an error on these tracks higher than that of the middle track. Consequently, the middle track better represents the TDMR device aging, and that is why our BER plots are for the middle track.

\section{Offline Learning}\label{offline}

In this section, we present our offline learning approach to optimally find the reconfiguration points as well as the relevant results. We consider three different optimality measures (setups of the objective function and the constraints), and we construct three different optimization problems according to these different optimality measures.

Through our error profile results, we observed that PIS patterns are dominant in the early lifetime of the TDMR device. When the device starts to age as the density increases, RTIS patterns become dominant (see \cite{ahh_otloco}). Consequently, our reconfiguration idea is as follows:
\begin{itemize}
	\item We start with the OP-LOCO coding scheme to eliminate PIS patterns when the device is fresh.
	\item Then, the coding scheme can be reconfigured to SP-LOCO in order to provide better performance.
	\item When RTIS patterns become dominant, the coding scheme can be again reconfigured to OT-LOCO.
	\item To provide better performance, ST-LOCO can be used as the last coding scheme via one more reconfiguration depending on the application needs.
\end{itemize}
Thus, we will assume that the reconfiguration order is OP-SP-OT-ST, where all of them are used unless the solutions of our optimization problems suggest otherwise.

\begin{remark}
Although the reconfiguration feature has been engineered to accommodate all coding schemes, optimal solution that finds the best compromise between maximizing storage capacity and minimizing complexity might lead to a scenario where one or more coding schemes remain unused.
\end{remark}


As mentioned earlier, in the following, we develop and solve three different optimization problems. Recall that for a LOCO code, the size of the adder that executes the rule governs the encoding-decoding complexity. In all of our optimization problems, reliable performance is characterized by a BER threshold. That is the reason why we obtain closed-form expressions for BER values in terms of the TD density via fitting the available data, which is the essence of offline learning in this work.

\begin{table}
\begin{center}
\caption{MSE Values of Different Function Models in Curve Fitting With Six Data Points for Middle-Track BER of OP-LOCO}
\label{tab_mse}
\begin{tabular}{|c|c|}
\hline
Function model & MSE \\ \hline
Exponential & $6.45\cdot10^{-7}$ \\ \hline
Gaussian & $3.59\cdot10^{-7}$ \\ \hline
Logarithmic & $3.89\cdot10^{-5}$ \\ \hline
Power & $3.90\cdot10^{-7}$ \\ \hline
Rational & $1.95\cdot10^{-4}$ \\ \hline
Sigmoidal & $2.84\cdot10^{-7}$ \\ \hline
Sinusoidal & $1.15\cdot10^{-6}$ \\ \hline
Third-degree Polynomial & $1.92\cdot10^{-7}$ \\ \hline
Fourth-degree Polynomial & $1.82\cdot10^{-7}$ \\ \hline
Seventh-degree Polynomial & $2.15\cdot10^{-23}$ \\ \hline
\end{tabular}
\end{center}
\vspace{-0.5em}
\end{table}

We fit on the error rate values obtained by simulating over the TDMR model, which are demonstrated in Fig.~\ref{fig_perf1}, to reach closed-form expressions for the error rate functions $f_{i}$, $i \in \{1,2,3,4\}$, where different $f_i$'s are for different coding schemes. In particular, these functions are for OP-LOCO, SP-LOCO, OT-LOCO, ST-LOCO coding schemes for $i = 1,2,3,4$, respectively. We collected data points within the TD density interval $[0.8, 1.5]$. Then, we fit using different function models and calculated the mean squared error (MSE) to find the most accurate form. It can be seen in the Table~\ref{tab_mse} that the seventh-degree polynomial gives the lowest MSE for the OP-LOCO coding scheme. MSE results for other coding schemes, skipped here for brevity, also indicate that seventh-degree polynomials give the lowest MSE. Thus, in the rest of the paper for offline learning, we use seventh-degree polynomials for curve fitting to represent the error rate functions. For online learning, as we shall see later, we use fifth-degree polynomials instead.

\subsection{First problem} \label{offline_first}

In this problem, the purpose is to maximize the storage capacity upon reconfiguring between the coding schemes while maintaining a reliable performance. The objective function of this optimization problem can be expressed as the weighted average of the rates of the coding schemes that are utilized. Weights here are the normalized lengths of the intervals in which each coding scheme utilized, and they form the optimization variable. This objective function directly indicates the percentage storage used for user information. The sum of these weights is equal to $1$ as they are normalized by the overall lifetime or TD density range. While switching between coding schemes, it is important to maintain a reliable and consistent error performance. We will be using BER to express the performance constraints in the optimization problem such that all of the coding schemes should have a BER under a predetermined value, leading to the optimization constraints. Both middle-track BER and all-tracks BER will be used in order to compare the results. In this problem, for middle-track and all-tracks BER constraints, the only difference will be the $f_i$ functions and associated thresholds since they will have different fitting data. However, all error functions will still have the seventh-degree polynomial form. Our indexing is always such that $1$ is for OP, $2$ is for SP, $3$ is for OT, and $4$ is for ST. Here, the rates are arranged in the vector $\mathbf{r}^\mathrm{T}=[r_1 \textup{ } r_2 \textup{ } r_3 \textup{ } r_4]$ and the weights (normalized intervals) are arranged in the vector $\mathbf{x}^\mathrm{T}=[x_1 \textup{ } x_2 \textup{ } x_3 \textup{ } x_4]$. Vectors are by default column vectors in this paper.

The first optimization problem is formulated as follows:
\begin{align}\label{main_problem}
\underset{\mathbf{x}}{\textrm{maximize}} \quad & \mathbf{r}^{\mathrm{T}}\mathbf{x} \\
\textrm{subject to} \quad & \mathbf{1}^{\mathrm{T}}\mathbf{x} = 1, \label{sum_one_const}\\
&x_i \geq 0, \, i=1,2,3,4, \label{big_zero_const}\\
&f_{1}(d_0 + x_1(d_1 - d_0)) \leq a, \label{fer_const1}\\
&f_{2}(d_0 + (x_1+x_2)(d_1 - d_0)) \leq a, \label{fer_const2}\\
&f_{3}(d_0 + (x_1+x_2+x_3)(d_1 - d_0)) \leq a, \label{fer_const3}\\
&f_{4}(d_1) \leq a.
\end{align}
In this optimization problem, $d_0$ represents the smallest density value we consider, which is $0.8$, and $d_1$ represents the largest density value we consider, which is $1.5$. The threshold $a$ represents the predetermined constant that middle-track BER or all-tracks BER is not allowed to surpass. Here, our optimization variable is $\mathbf{x}$, which contains the fraction of density durations for various codes as discussed above. That is why the entries of $\mathbf{x}$ sum up to $1$ as indicated by the constraint (\ref{sum_one_const}). The constraint indicated by (\ref{big_zero_const}) is obvious since each $x_i$ represents a fraction of density duration for the respective coding scheme. Constraints (\ref{fer_const1}), (\ref{fer_const2}), and (\ref{fer_const3}) represent the performance constraints we impose, and functions $f_1$, $f_2$, $f_3$, and $f_4$ are the closed-form expressions of middle-track BER or all-tracks BER with input arguments as the relevant TD density values. $f_{4}(d_1) \leq a$ is the immediate constraint that does not depend on the optimization variable because of the nature of this problem. One can use minimization of the negative of the objective function instead of maximization of the objective function above as well. Since we use seventh-degree polynomials to represent the BER values, we represent the performance functions obtained by curve fitting as follows:
\vspace{-0.7em}\begin{equation}
\label{curve_fit_main}
f_i(x) = \sum_{j=1}^8 m_{i,j}x^{8-j},
\end{equation}
where $m_{i,j}$ values are given in Table~\ref{fitting_coeff}. Moreover, the indexing is as specified above.

\begin{remark}
Due to the nature of the optimization problem, solution does not require an advanced neural network or a machine learning model. Curve fitting is used to obtain the closed-form expressions of the BER functions in the constraints, and optimization methods are then used to solve the problem. In the online setting, the storage device controller should be able to execute the procedure of collecting data, fitting, and optimization to decide on code reconfiguration.
\end{remark}

\begin{table}
\begin{center}
\caption{Coefficients of the Seventh-Degree Polynomials Resulting From Curve Fitting}
\label{fitting_coeff}
\begin{tabular}{|c|c|c|c|c|}
\hline
\multicolumn{1}{|c|}{} & \multicolumn{1}{c|}{\textbf{$m_{i,1}$}} & \multicolumn{1}{c|}{\textbf{$m_{i,2}$}} & \multicolumn{1}{c|}{\textbf{$m_{i,3}$}} & \multicolumn{1}{c|}{\textbf{$m_{i,4}$}} \\ \hline
\textbf{OP-LOCO}       & $-0.100$                            & $2.082$                             & $-10.710$                             & $25.300$                              \\ \hline
\textbf{SP-LOCO}       & $-0.516$                            & $3.952$                             & $-12.800$                            & $22.660$                             \\ \hline
\textbf{OT-LOCO}       & $-0.009$                            & $0.076$                             & $-0.252$                            & $0.446$                             \\ \hline
\textbf{ST-LOCO}       & $0.112$                           & $-0.837$                             & $2.486$                             & $-4.048$                             \\ \hline
                       & \multicolumn{1}{c|}{\textbf{$m_{i,5}$}} & \multicolumn{1}{c|}{\textbf{$m_{i,6}$}}  & \multicolumn{1}{c|}{\textbf{$m_{i,7}$}}  & \multicolumn{1}{c|}{\textbf{$m_{i,8}$}}     \\ \hline
\textbf{OP-LOCO}       & $-32.100$                             & $22.620$                              & $-8.348$                             & $1.258$                             \\ \hline
\textbf{SP-LOCO}       & $-23.610$                            & $14.460$                               & $-4.807$                            & $0.670$                             \\ \hline
\textbf{OT-LOCO}       & $-0.462$                            & $0.281$                             & $-0.093$                            & $0.013$                             \\ \hline
\textbf{ST-LOCO}       & $3.901$                            & $-2.222$                             & $0.6925$                            & $-0.091$                              \\ \hline
\end{tabular}
\end{center}
\end{table}

Now, we show that our problem is convex and it is a linear programming problem. Even though the constraints of the optimization problem contain seventh-degree polynomials, which do not result in convex feasible sets in general, due to the restricted region of interest, the feasible set of our optimization problem is convex. The following discussion is for $a=10^{-3}$, middle-track BER constraints, and the first problem. Having said that, the conclusion (the problem being convex and a linear program) is the same for all other setups we work on. The feasible set of the optimization problem, obtained via the original constraints, is as follows:
\begin{align}
x_1+x_2+x_3+x_4=1, \label{eqn_lp1} \\
x_1\geq0, \textup{ } x_2\geq0, \textup{ } x_3\geq0, \textup{ } x_4\geq0, \label{eqn_lp2} \\
x_1\leq0.2624, \label{eqn_lp3} \\
x_1+x_2\leq0.4082, \label{eqn_lp4} \\
x_1+x_2+x_3\leq0.9121. \label{eqn_lp5}
\end{align}
Thus, the feasible set is the intersection of one hyperplane and multiple half-spaces. Since intersection of convex sets is convex, the feasible set of the optimization problem is also convex. Furthermore, the objective function of the optimization problem is a linear function (convex and concave). Therefore, we conclude that the optimization problem is convex and is a linear programming problem. Hence, convex optimization methods can be used to solve the problem. Specifically, we use Karush-Kuhn-Tucker (KKT) conditions.

Next, we start by forming the Lagrangian of the optimization problem:
\begin{align}
&L(\mathbf{x}, \boldsymbol{\lambda}, \nu) = -\mathbf{r}^{\mathrm{T}}\mathbf{x}-\lambda_1x_1-\lambda_2x_2-\lambda_3x_3-\lambda_4x_4 \nonumber \\
&+ \lambda_5\Bigg[\sum_{j=1}^8 m_{1,j}(d_0+x_1(d_1-d_0))^{8-j} - a\Bigg] \nonumber \\
&+ \lambda_6\Bigg[\sum_{j=1}^8 m_{2,j}(d_0+(x_1+x_2)(d_1-d_0))^{8-j} - a\Bigg] \nonumber \\
&+ \lambda_7\Bigg[\sum_{j=1}^8 m_{3,j}(d_0+(x_1+x_2+x_3)(d_1-d_0))^{8-j} - a\Bigg] \nonumber \\
&+ \nu(x_1+x_2+x_3+x_4-1),
\end{align}
where $\boldsymbol{\lambda}^\textrm{T} = [\lambda_1 \textup{ } \lambda_2 \textup{ } \lambda_3 \textup{ } \lambda_4 \textup{ } \lambda_5 \textup{ } \lambda_6 \textup{ } \lambda_7]$. For the simplicity of reading, we will be using the following auxiliary functions:
\vspace{-0.1em}\begin{align}
g_1(x_1) &= (d_1-d_0)\sum_{j=1}^{7} \big [ (8-j)m_{1,j}(d_0 \nonumber \\
&+ x_1(d_1-d_0))^{7-j} \big ], \label{eqn_g1} \\
g_2(x_1,x_2) &= (d_1-d_0)\sum_{j=1}^{7} \big [ (8-j)m_{2,j}(d_0 \nonumber \\
&+ (x_1+x_2)(d_1-d_0))^{7-j} \big ], \label{eqn_g2} \\
g_3(x_1,x_2,x_3) &= (d_1-d_0)\sum_{j=1}^{7} \big [ (8-j)m_{3,j}(d_0 \nonumber \\
&+ (x_1+x_2+x_3)(d_1-d_0))^{7-j} \big ] \label{eqn_g3}.
\end{align}
We need to obtain the partial derivatives of the Lagrangian with respect to $x_i$, for all $i$, to apply the vanishing gradient KKT condition:
\begin{align}
\frac{\partial L}{\partial x_1} &= [(\lambda_5g_1(x_1)) + (\lambda_6g_2(x_1,x_2))  \nonumber \\
&+ (\lambda_7g_3(x_1,x_2,x_3))] - r_1 - \lambda_1 + \nu, \\
\frac{\partial L}{\partial x_2} &= [(\lambda_6g_2(x_1,x_2)) + (\lambda_7g_3(x_1,x_2,x_3))] \nonumber \\
&- r_2 - \lambda_2 + \nu, \\
\frac{\partial L}{\partial x_3} &= [\lambda_7g_3(x_1,x_2,x_3)] - r_3 - \lambda_3 + \nu, \\
\frac{\partial L}{\partial x_4} &= -r_4 - \lambda_4 + \nu.
\end{align}
The vanishing gradient condition is that $\frac{\partial L}{\partial x_i} = 0$, for all $i$. Thus, we obtain the following set of equations:
\begin{align}
\lambda_1 &= -r_1 + \nu + \lambda_5g_1(x_1) + \lambda_6g_2(x_1,x_2)  \nonumber \\
&+  \lambda_7g_3(x_1,x_2,x_3), \label{lamb1} \\
\lambda_2 &= -r_2 + \nu + \lambda_6g_2(x_1,x_2) + \lambda_7g_3(x_1,x_2,x_3), \label{lamb2} \\
\lambda_3 &= -r_3 + \nu + \lambda_7g_3(x_1,x_2,x_3), \label{lamb3} \\
\lambda_4 &= -r_4 + \nu. \label{lamb4}
\end{align}
Then, by using (\ref{lamb1})-(\ref{lamb4}), we obtain $\lambda_5$, $\lambda_6$, and $\lambda_7$ as follows:
\begin{equation}\label{lamb5}
\lambda_5 = \frac{\lambda_1 - \lambda_2 + r_1 - r_2}{g_1(x_1)},
\end{equation}
\begin{equation}\label{lamb6}
\lambda_6 = \frac{\lambda_2 - \lambda_3 + r_2 - r_3}{g_2(x_1,x_2)},
\end{equation}
\begin{equation}\label{lamb7}
\lambda_7 = \frac{\lambda_3 + r_3 - \nu}{g_3(x_1,x_2,x_3)}.
\end{equation}
The solution is then obtained via combining the above equations with the complementary slackness and the dual constraint KKT conditions, where the latter is $\boldsymbol{\lambda} \succeq 0$. The solution will be investigated in Subsection~\ref{offline_second} in detail because of the similarity of the two problems. Having said that, the conceptual essence of the solution is stated in Remarks~\ref{rem_sol1} and \ref{rem_sol2}.

\begin{remark}\label{rem_sol1}
The solution can be derived aided by the principles of complementary slackness as well as the dual constraint. To satisfy both, it is necessary that $f_i(\cdot)-a=0$, for all $i$ (this will be shown in Subsection~\ref{offline_second}). This condition results in a logically-expected solution; the selected, higher-rate coding scheme should be utilized until its BER performance reaches the threshold $a$, at which point the system reconfigures to a new, lower-rate coding scheme. Repeating this process three times gives the solution to the first optimization problem. Despite its simplicity, this solution is crucial for addressing subsequent challenges in our analysis.
\end{remark}

\begin{remark}\label{rem_sol2}
If we replace the BER constraints (\ref{fer_const1}), (\ref{fer_const2}), and (\ref{fer_const3}) by the linear constraints (\ref{eqn_lp3}), (\ref{eqn_lp4}), and (\ref{eqn_lp5}), we can see that the solution of the first problem can be obtained directly from the finite corner of the open polytope characterized by (\ref{eqn_lp1}), (\ref{eqn_lp2}), (\ref{eqn_lp3}), (\ref{eqn_lp4}), and (\ref{eqn_lp5}). That is, $\mathbf{x}^\mathrm{T} = [0.2624 \textup{ } 0.1458 \, \allowbreak 0.5039 \textup{ } 0.0879]$. However, the aforementioned analysis is still very important for the following two, more advanced problems.
\end{remark}

\subsection{Second problem} \label{offline_second}

In this problem, we aim to find the best compromise between maximizing storage capacity and minimizing complexity. Therefore, the objective function of the optimization problem is expressed as the weighted average of the rates of the coding schemes utilized minus the weighted average of the adder sizes of the coding schemes with a fractional scaling constant so that the complexity term does not dominate the objective function. The adder sizes are arranged in the vector $\mathbf{b}^\mathrm{T}=[b_1 \textup{ } b_2 \textup{ } b_3 \textup{ } b_4]$ and the weights (normalized intervals) are arranged in the vector $\mathbf{x}$, which is the optimization variable.

It is important to note that there is only one adder being used throughout the lifetime of the device, and it has the maximum size needed. This adder is the one being reconfigured at the switching points. Whenever we say adder sizes, we mean how many $1$-bit full adders within that single adder are used for different LOCO coding schemes. Thus, complexity throughout the paper means execution-time complexity.

The second optimization problem is formulated as follows:
\begin{align}
\underset{\mathbf{x}}{\textrm{maximize}} \quad & \mathbf{r}^{\mathrm{T}}\mathbf{x}-\frac{1}{c}\mathbf{b}^{\mathrm{T}}\mathbf{x} \label{second_problem_long_objective} \\
\textrm{subject to} \quad & \mathbf{1}^{\mathrm{T}}\mathbf{x} = 1,\\
&x_i \geq 0, \, i=1,2,3,4, \\
&f_{1}(d_0 + x_1(d_1 - d_0)) \leq a,\\
&f_{2}(d_0 + (x_1+x_2)(d_1 - d_0)) \leq a,\\
&f_{3}(d_0 + (x_1+x_2+x_3)(d_1 - d_0)) \leq a,\\
&f_{4}(d_1) \leq a.
\end{align}
In this problem, $f_i(x)$ functions are identical to those in the first optimization problem, with the same coefficients. One can transform the optimization problem to the following one:
\begin{align}\label{main_problem}
\underset{\mathbf{x}}{\textrm{minimize}} \quad & -\mathbf{k}^{\mathrm{T}}\mathbf{x} \\
\textrm{subject to} \quad & \mathbf{1}^{\mathrm{T}}\mathbf{x} = 1,\\
&x_i \geq 0, \, i=1,2,3,4, \\
&f_{1}(d_0 + x_1(d_1 - d_0)) \leq a,\\
&f_{2}(d_0 + (x_1+x_2)(d_1 - d_0)) \leq a,\\
&f_{3}(d_0 + (x_1+x_2+x_3)(d_1 - d_0)) \leq a,\\
&f_{4}(d_1) \leq a,
\end{align}
where $\mathbf{k}^{\mathrm{T}}=\mathbf{r}^{\mathrm{T}}-\frac{1}{c}\mathbf{b}^{\mathrm{T}}$. Now, instead of maximizing the objective function, we minimize the negative of it. Here, $c$ is used as a scaling constant and it will affect the solution. Observe that the problems investigated in Subsection~\ref{offline_first} and Subsection~\ref{offline_second} are similar. However, their solutions will not be identical as a result of the difference between $\mathbf{r}$ and $\mathbf{k}$. Here, the relations between $k_i$ values dictate which coding schemes to be used. As we did for the previous problem, we can show that the second problem is also convex. Thus, we can use KKT conditions to obtain the optimal solution. In Subsection~\ref{offline_first}, we obtained Equations (\ref{lamb5}), (\ref{lamb6}), and (\ref{lamb7}) for the first problem. Because of the similarity of the two problems where the only difference is that $\mathbf{r}$ is replaced with $\mathbf{k}$ in the second one, we can obtain the following set of equations in a way similar to that of the first optimization problem:
\begin{equation}\label{second_lamb5}
\lambda_5 = \frac{\lambda_1 - \lambda_2 + k_1 - k_2}{g_1(x_1)},
\end{equation}
\begin{equation}\label{second_lamb6}
\lambda_6 = \frac{\lambda_2 - \lambda_3 + k_2 - k_3}{g_2(x_1,x_2)},
\end{equation}
\begin{equation}\label{second_lamb7}
\lambda_7 = \frac{\lambda_3 + k_3 - \nu}{g_3(x_1,x_2,x_3)}.
\end{equation} \newline
Observe that Equations (\ref{lamb1})-(\ref{lamb4}) for $\lambda_1$, $\lambda_2$, $\lambda_3$, and $\lambda_4$ remain the same, but with each $r_i$ being replaced with the relevant $k_i$, for all $i$ in $\{1,2,3,4\}$. A detailed analysis for the rest of the solution is shown in Remark~\ref{Rmk_prob2Soln}. Since the complete solution involves a lengthy case-by-case analysis, we only discuss certain cases in the remark to illustrate the logic for simplicity and brevity.

\begin{table}
\begin{center}
\caption{Rates and Adder Sizes of All Coding Schemes at Length $23$}
\label{tab_params}
\begin{tabular}{|c|c|c|c|c|}
\hline
           & OP-LOCO & SP-LOCO & OT-LOCO & ST-LOCO \\ \hline
Rate       & $0.9306$  & $0.8800$  & $0.8267$  & $0.7436$  \\ \hline
Adder size & $67$      & $17$      & $59$      & $30$      \\ \hline
\end{tabular}
\end{center}
\vspace{-0.5em}
\end{table}

\begin{table}
\begin{center}
\caption{Solution of the Second Optimization Problem With Middle-Track Bit Error Rate Constraint at Most $a=10^{-3}$}
\label{tab_opt1}
\vspace{-1.0em}
\begin{tabular}{|c|c|c|c|c|c|}
\hline
& $x_1$ & $x_2$ & $x_3$ & $x_4$ & Capacity \\ \hline
\multicolumn{6}{|c|}{$c>988$} \\ \hline
\makecell{%
$x_1\neq0, x_2\neq0$,\\
$x_3\neq0, x_4\neq0$
} & $0.2624$ & $0.1458$ & $0.5039$ & $0.0879$ & $0.8544$ \\ \hline 
$x_1=0$ & $0$ & $0.4082$ & $0.5039$ & $0.0879$ & $0.8412$ \\ \hline
$x_1=0, x_2=0$ & $0$ & $0$ & $0.9121$ & $0.0879$ & $0.8194$ \\ \hline
$x_1=0, x_3=0$ & $0$ & $0.4082$ & $0$ & $0.5918$ & $0.7993$ \\ \hline
$x_2=0, x_3=0$ & $0.2624$ & $0$ & $0$ & $0.7376$ & $0.7927$ \\ \hline
\multicolumn{6}{|c|}{$349<c<988$} \\ \hline
$x_1=0$ & $0$ & $0.4082$ & $0.5039$ & $0.0879$ & $0.8412$ \\ \hline
$x_1=0, x_2=0$ & $0$ & $0$ & $0.9121$ & $0.0879$ & $0.8194$ \\ \hline
$x_1=0, x_3=0$ & $0$ & $0.4082$ & $0$ & $0.5918$ & $0.7993$ \\ \hline
\multicolumn{6}{|c|}{$0<c<349$} \\ \hline
$x_1=0, x_3=0$ & $0$ & $0.4082$ & $0$ & $0.5918$ & $0.7993$ \\ \hline
\end{tabular}
\end{center}
\vspace{-0.5em}
\end{table}

The rates and adder sizes of all four coding schemes are listed in Table~\ref{tab_params}, and the following analysis depends on them. Our results are trivially generalizable to any set of parameters. The value of $c$ can be specified according to the storage designer priorities. We will be investigating different ranges of $c$ because $c$ value alters the relations between $k_i$'s as follows:
\vspace{-0.1em}\begin{align}
c>988 &\implies k_1>k_2, \, k_2>k_3, \, k_3>k_4, \label{ki_ineq1} \\
349<c<988 &\implies k_1<k_2, \, k_2>k_3, \, k_3>k_4, \label{ki_ineq2} \\
0<c<349 &\implies k_1<k_2, \, k_2>k_3, \, k_2>k_4, \, k_3<k_4. \label{ki_ineq3}
\end{align}
Table~\ref{tab_opt1} shows the different cases where there exists an optimal solution for different ranges of $c$ values. Results in Table~\ref{tab_opt1} are obtained by setting the middle-track BER threshold to $a = 10^{-3}$, which can also be altered if needed.

We focus now on the insights of the solutions. Here, it is important to note that $k_i$ values matter while obtaining the optimal solution. When the order of $k_i$ values changes, it affects which coding schemes to be used. If we focus on the region $349<c<988$, one can confirm that $k_1<k_2$. This inequality forces us not to use the OP-LOCO coding scheme because OP-LOCO has the worst error performance, i.e., starting with SP-LOCO is better for error performance, and OP-LOCO also has the largest adder size compared with the other coding schemes. In other words, using OP-LOCO does not offer any advantage when $k_1<k_2$, and instead, we should start with the SP-LOCO coding scheme. Another interpretation is that, for the region $349<c<988$, reallocating the TD density share from OP-LOCO to SP-LOCO improves the error performance (note that this satisfies the constraints trivially) and further reduces the objective function \ref{main_problem}. Thus, the feasible set consists of the set of relevant $\mathbf{x}$ values where $x_1=0$. Similarly, in the region $0<c<349$, $k_1<k_2$ and $k_3<k_4$. Thus, the feasible set consists of the set of relevant $\mathbf{x}$ values where $x_1=0$ and $x_3=0$. On the other hand, in the region $c>988$, all $k_i$'s are ordered descendingly with $i$, which means there is no additional restriction on the usage of any of the coding schemes. Therefore, in the region $c>988$, the solution of this problem becomes identical to the solution of the first problem due to the ordered $k_i$'s, which resemble the ordered $r_i$'s. This is also expected because very large $c$ values eliminate the effect of $\mathbf{b}^{\mathrm{T}}\mathbf{x}$ in (\ref{second_problem_long_objective}) on the solution, and the two problems become identical then. These insights, in addition to the mathematical derivations, are used to obtain the solutions given in Table~\ref{tab_opt1} and Table~\ref{tab_opt1_all}. As shown in Remark~\ref{Rmk_prob2Soln}, linear programming basics can be used as well.

\begin{remark}\label{Rmk_prob2Soln}
We demonstrate how the results are derived from (\ref{second_lamb5})-(\ref{second_lamb7}) and mathematically explain why the condition $k_1<k_2$ prevents the use of the OP-LOCO coding scheme in the region $349<c<988$. The solution is obtained by examining all possible combinations of $\lambda_1$, $\lambda_2$, $\lambda_3$, and $\lambda_4$, where each can be either $0$ or strictly greater than $0$.
\begin{enumerate}
	\item First, we assume that $\lambda_1 = 0$, $\lambda_2 = 0$, $\lambda_3 = 0$, and $\lambda_4 = 0$. Then, Equation~(\ref{second_lamb5}) turns into
\vspace{-0.1em}\begin{equation}\label{remark_lamb5}
\lambda_5 = \frac{k_1 - k_2}{g_1(x_1)}.
\end{equation}
Now, since $k_1 < k_2$, the numerator of the right-hand side in (\ref{remark_lamb5}) is negative. Then, by the dual constraints, the denominator of the right-hand side in (\ref{remark_lamb5}), $g_1(x_1)$, should be negative so that $\lambda_5$ becomes positive as it should be. The complementary slackness condition associated with $\lambda_5$ is $\lambda_5(f_{1}(d_0 + x_1(d_1 - d_0))-a)=0$. It is obvious that $\lambda_5\neq0$ since $k_1\neq~k_2$. Therefore, $f_{1}(d_0 + x_1(d_1 - d_0))-a=0$. The value of $x_1$ that satisfies $f_{1}(d_0 + x_1(d_1 - d_0))-a=0$ is $x_1=0.2624$. However, $x_1=0.2624$ does not make $g_1(x_1)$ in (\ref{eqn_g1}) negative, which is a contradiction. Thus, we conclude that $\lambda_i=0$ for $i \in \{1,2,3,4\}$ is not possible.
	\item Now, we assume that $\lambda_1\neq0$ and $\lambda_2 = \lambda_3 = \lambda_4 = 0$. This implies that $x_1=0$ by complementary slackness. Equation~(\ref{second_lamb5}) then turns into
\begin{equation}\label{remark_lamb5_2}
\lambda_5 = \frac{\lambda_1 + k_1 - k_2}{g_1(x_1=0)}.
\end{equation}
Since $x_1=0$, by direct substitution, $f_{1}(d_0 + x_1(d_1 - d_0))-a = f_{1}(d_0)-a <0$. Then, by complementary slackness, $\lambda_5=0$. The complementary slackness conditions associated with $\lambda_6$ and $\lambda_7$ for $x_1=0$ are $\lambda_6(f_{2}(d_0 + x_2(d_1 - d_0))-a)=0$ and $\lambda_7(f_{3}(d_0 + (x_2+x_3)(d_1 - d_0))-a)=0$, respectively. From (\ref{lamb4}), $\lambda_4 = -k_4 + \nu = 0$, leading to $\nu = k_4$. Using these in (\ref{second_lamb6}) and (\ref{second_lamb7}), we reach:
\vspace{-0.1em}\begin{align}
\lambda_6&=\frac{k_2-k_3}{g_2(x_1=0,x_2)}, \label{lamb6_ex1}\\
\lambda_7&=\frac{k_3-k_4}{g_3(x_1=0,x_2,x_3)}. \label{lamb7_ex1}
\end{align}
Now, since $k_2>k_3$ and $k_3>k_4$, $\lambda_6 \neq 0$ and $\lambda_7\neq0$. Then, by complementary slackness and that $x_1=0$, $f_{2}(d_0 + x_2(d_1 - d_0))-a=0$ and $f_{3}(d_0 + (x_2+x_3)(d_1 - d_0))-a=0$ must be satisfied. As a result, $x_2=0.4082$, $x_3=0.5039$, and $x_4=0.0879$.
	\item By investigating every possible configuration in a way similar to that of the previous cases, we can see that any possible configuration must contain the condition $\lambda_1\neq0$ for $349<c<988$.
	\item Once we reach the condition that $\lambda_1\neq0$, leading to $x_1=0$, we can also solve the problem using linear programming basics. In particular, the feasible set represented by the open polytope in Remark~\ref{rem_sol2} is updated such that only the intersection with the hyperplane $x_1=0$ is kept in the set. Thus, the new finite corner leads to the solution $\mathbf{x}^\mathrm{T} = [0 \textup{ } 0.4082 \textup{ } 0.5039 \textup{ } 0.0879]$.
\end{enumerate}
Here, we investigated the region $349<c<988$. The complete solution can be obtained similarly for all other regions.
\end{remark}

\begin{remark}
We skipped the equality cases for the inequalities given in (\ref{ki_ineq1})-(\ref{ki_ineq3}) for brevity and due to the fact that equalities are satisfied only for specific values of $c$. For completeness, we also introduce the equality cases in this remark.
\begin{align}
c\geq988 &\implies k_1\geq k_2, \, k_2>k_3, \, k_3>k_4, \label{ki_eq1} \\
349\leq c<988 &\implies k_1<k_2, \, k_2>k_3, \, k_3\geq k_4, \label{ki_eq2} \\
0<c<349 &\implies k_1<k_2, \, k_2>k_3, \, k_2>k_4, \, k_3<k_4. \label{ki_eq3}
\end{align}
Since we examined $349<c<988$ in Remark~\ref{Rmk_prob2Soln}, we again discuss the same region here but with $349\leq c<988$. For the specific case of $c=349$, $k_3=k_4$. Therefore, complementary slackness for the optimal solution will be achieved via both $\lambda_7=0$ as well as $f_{3}(d_0 + (x_2+x_3)(d_1 - d_0))-a=0$, which implies no impact on the results discussed in Remark~\ref{Rmk_prob2Soln}. The same logic also applies for the case of $c=988$.
\end{remark}

Next, we also investigate the optimization trade-off with all-tracks BER constraints. Thus, we change the constraints in the optimization problem to use all-tracks BER fitting functions and threshold. Seventh-degree polynomials still give the lowest MSE. Here, we set the BER threshold as $a=10^{-2}$. Table~\ref{tab_opt1_all} shows the solutions of the optimization problem with all-tracks BER performance constraints.

\begin{table}
\begin{center}
\caption{Solution Of The Second Optimization Problem With All-Tracks Bit Error Rate Constraint at Most $a=10^{-2}$}
\label{tab_opt1_all}
\begin{tabular}{|c|c|c|c|c|c|}
\hline
& $x_1$ & $x_2$ & $x_3$ & $x_4$ & Capacity \\ \hline
\multicolumn{6}{|c|}{$c>988$} \\ \hline
\makecell{%
$x_1\neq0, x_2\neq0$,\\
$x_3\neq0, x_4\neq0$
} & $0.8113$ & $0.0930$ & $0.0957$ & $0$ & $0.9160$ \\ \hline
$x_1=0$ & $0$ & $0.9043$ & $0.0957$ & $0$ & $0.8749$ \\ \hline
$x_1=0, x_2=0$ & $0$ & $0$ & $1$ & $0$ & $0.8267$ \\ \hline
$x_1=0, x_3=0$ & $0$ & $0.9043$ & $0$ & $0.0957$ & $0.8669$ \\ \hline
$x_2=0, x_3=0$ & $0.8113$ & $0$ & $0$ & $0.1887$ & $0.8953$ \\ \hline
\multicolumn{6}{|c|}{$349<c<988$} \\ \hline
$x_1=0$ & $0$ & $0.9043$ & $0.0957$ & $0$ & $0.8749$ \\ \hline
$x_1=0, x_2=0$ & $0$ & $0$ & $1$ & $0$ & $0.8267$ \\ \hline
$x_1=0, x_3=0$ & $0$ & $0.9043$ & $0$ & $0.0957$ & $0.8669$ \\ \hline
\multicolumn{6}{|c|}{$0<c<349$} \\ \hline
$x_1=0, x_3=0$ & $0$ & $0.9043$ & $0$ & $0.0957$ & $0.8669$ \\ \hline
\end{tabular}
\end{center}
\end{table}

\begin{remark}
By observing the difference between the middle-track and all-tracks results, we can deduce that the middle-track constraint typically allows a longer lifetime at a lower average storage capacity, while the all-tracks constraint typically allows a shorter lifetime, if the middle-track BER threshold is used, at a higher average storage capacity. These conclusions may change if the BER thresholds change.
\end{remark}

\subsection{Third Problem}

In this problem, we add an additional constraint to the first problem. In particular, the average adder size should not exceed a predetermined threshold $z$. Thus, the third optimization problem is formulated as follows:
\begin{align}\label{third_problem}
\underset{\mathbf{x}}{\textrm{minimize}} \quad & -\mathbf{r}^{\mathrm{T}}\mathbf{x} \\
\textrm{subject to} \quad & \mathbf{1}^{\mathrm{T}}\mathbf{x} = 1, \\
&x_i \geq 0, \, i=1,2,3,4, \\
&f_{1}(d_0 + x_1(d_1 - d_0)) \leq a, \\
&f_{2}(d_0 + (x_1+x_2)(d_1 - d_0)) \leq a, \\
&f_{3}(d_0 + (x_1+x_2+x_3)(d_1 - d_0)) \leq a, \\
&f_{4}(d_1) \leq a, \\
&\mathbf{b}^{\mathrm{T}}\mathbf{x} \leq z. \label{third_comp_const}
\end{align}
In this problem, $f_i(x)$ functions are identical to those in the first optimization problem, with the same coefficients. Similar to the previous problems, this problem is also convex and it is another linear programming problem. Here, we will attempt solving the problem via linear programming basics. In particular, we will investigate the corners resulting from updating the open polytope in Remark~\ref{rem_sol2} such that only the intersection with the half space $\mathbf{b}^{\mathrm{T}}\mathbf{x} \leq z$ is kept in the set.

\begin{algorithm}[H]
\caption{Vertex (Corner Point) Finding and LP Optimal Vertex Specification}
\label{alg:vertex_enum_lp}
\begin{algorithmic}[1]

\Require Matrix $\mathbf{A}\in\mathbb{R}^{m\times n}$ and vector $\mathbf{b}\in\mathbb{R}^m$ such that they define the inequality constraints $\mathbf{Ax}\leq \mathbf{b}$, $\mathbf{x}\in\mathbb{R}^n$.
\Require Equality constraints $\mathbf{A}_{\textup{eq}}\mathbf{x} = \mathbf{b}_{\textup{eq}}$, where $\mathbf{A}_\textup{eq}\in\mathbb{R}^{p\times n}$ and $\mathbf{b_\textup{eq}}\in\mathbb{R}^p$.
\Require Rate vector $\mathbf{r}\in\mathbb{R}^n$.

\State $r \gets \mathrm{rank}(\mathbf{A}_{\textup{eq}})$.
\State $d \gets n - r$.
\State Initialize $\mathcal{V} \gets \varnothing$.

\ForAll{index sets $S \subseteq \{1,\ldots,m\}$ with $|S| = d$}
    \State Form $\mathbf{A}_S$ from rows of $\mathbf{A}$ indexed by $S$.
    \State Form $\mathbf{b}_S$ from the entries in $\mathbf{b}$ indexed by $S$.
    \State $\mathbf{M} \gets \begin{bmatrix} \mathbf{A}_{\textup{eq}} \\ \mathbf{A}_S \end{bmatrix}\).
    \State $\mathbf{\boldsymbol{\rho}} \gets \begin{bmatrix} \mathbf{b}_{\textup{eq}} \\ \mathbf{b}_S \end{bmatrix}\).

    \If{ $\mathrm{rank}(\mathbf{M}) < n$ }
        \State \textbf{go to} Step 4 to select a new $S$.
    \EndIf

    \State Solve $\mathbf{M x} = \boldsymbol{\rho}$ for $\mathbf{x}$.

    \If{ ($\mathbf{A}_{\textup{eq}}\mathbf{x} = \mathbf{b}_{\textup{eq}}$)
          $\land$ ($\mathbf{Ax} \le \mathbf{b}$) }
        \State $\mathcal{V} \gets \mathcal{V} \cup \{\mathbf{x}\}$.
    \EndIf

\EndFor

\State Remove duplicates from $\mathcal{V}$ (if any).

\If{$\mathcal{V} = \varnothing$}
    \State \Return ``No vertices found (polytope may be empty).''
\EndIf

\State $\mathbf{x}^\star \gets \arg\min_{\mathbf{x}\in\mathcal{V}} \, -\mathbf{r}^{\mathrm{T}} \mathbf{x}$.
\State $f^\star \gets -\mathbf{r}^{\mathrm{T}} \mathbf{x}^\star$.

\State \Return Vertices \(\mathcal{V}\), optimal vertex $\mathbf{x}^\star$, and minimum objective function value $f^\star$.

\end{algorithmic}
\end{algorithm}

The feasible set here is as follows:
\begin{align}
x_1+x_2+x_3+x_4=1, \label{eqn_new_lp1}\\
x_1\geq0, \textup{ } x_2\geq0, \textup{ } x_3\geq0, \textup{ } x_4\geq0, \label{eqn_new_lp2}\\
x_1\leq0.2624, \label{eqn_new_lp3}\\
x_1+x_2\leq0.4082, \label{eqn_new_lp4}\\
x_1+x_2+x_3\leq0.9121, \label{eqn_new_lp5}\\
\mathbf{b}^{\mathrm{T}}\mathbf{x} \leq z. \label{eqn_new_lp6}
\end{align}
Algorithm~\ref{alg:vertex_enum_lp} finds the vertices (corners) of the feasible set for any value of $z$, which is part of the inequality constraints, once we list the inequality and equality constraints as inputs, and it also specifies the optimal value of $\mathbf{x}$. We now briefly analyze the problem according to the regions dictated by $z$:
\begin{itemize}
\item First, we find the $z$ value below which there does not exist a solution. This $z$ value is the minimum value such that the half space $\mathbf{b}^{\mathrm{T}}\mathbf{x} \leq z$ intersects with the open polytope given in the original feasible set (\ref{eqn_new_lp1})-(\ref{eqn_new_lp5}). To find this $z$ value, we need to utilize the coding scheme with the smallest adder size, which is SP-LOCO with adder size $17$, as much as possible. Thus, $x_2$ should be maximized. Since $x_1+x_2\leq0.4082$, $x_2=0.4082$. The remaining share should either go to $x_3$ or $x_4$. Since we are searching for the minimum $z$, the remaining share should go the coding scheme with a smaller adder size, which is ST-LOCO. Consequently, $x_4=1-x_2=0.5918$, and $x_1=x_3=0$. Therefore, $\mathbf{b}^{\mathrm{T}}\mathbf{x}=24.6934$, and for $z<24.6934$, no feasible solution exists.

\item Next, we investigate higher $z$ values and try to improve the objective function while characterizing the next region. The question is: with a higher budget, how can we improve the objective function as much as possible? First, taking from the share of $x_2$ and giving to either $x_3$ or $x_4$ is not logical because $x_2$ has higher rate and lower adder size. Possible options are giving from $x_2$ to $x_1$ or from $x_4$ to $x_3$. We compare the improvements in the objective function that we will get by increasing the average adder size. For the option $x_2$ to $x_1$, it is $(0.9306-0.88)/(67-17)=0.001012$. For the option $x_4$ to $x_3$, the improvement in the objective function per unit increase in the adder size is $(0.8267-0.7436)/(59-30)=0.0028655$. Therefore, we should pick the second option in this region, and the first option will eventually be employed in the next region. Consequently,
\begin{align}
&x_1+x_2+x_3\leq0.9121 \implies \nonumber \\ 
&x_3\leq0.9121-0.4082=0.5039.
\end{align}
When $x_3=0.5039$, $x_4=0.0879$ and $\mathbf{b}^{\mathrm{T}}\mathbf{x}=39.3065$. Now, we find the optimal $\mathbf{x}$ in this region, $24.6934\leq z\leq39.3065$. Moreover,
\begin{align}
z &= \mathbf{b}^{\mathrm{T}}\mathbf{x}= 17\times0.4082+59\times x_3 \nonumber \\
&+30\times(0.5918-x_3) \implies x_3=\frac{z-24.6934}{29}, \nonumber \\
x_4 &= 0.5918-x_3 = \frac{41.8556-z}{29}.
\end{align}
Thus, for $24.6934\leq z\leq39.3065$, optimal solution is
\begin{align}
x_1 &= 0, \textup{ } x_2 = 0.4082, \nonumber \\
x_3 &= \frac{z-24.6934}{29}, \textup{ } x_4 = \frac{41.8556-z}{29}.
\end{align}

\item At the end of the previous region (highest $z$), we have $\mathbf{x}^\mathrm{T}=[0 \textup{ } 0.4082 \textup{ } 0.5039 \textup{ } 0.0879]^\mathrm{T}$. To specify the next region, we again investigate the answer of the question: with a higher $z$ value, how can we improve the objective function as much as possible? Notice that $x_4$ is already at its minimum value due to the constraints (\ref{eqn_new_lp1}) and (\ref{eqn_new_lp5}). The only valid option then that satisfies the constraints and improves the objective function is that we should give from $x_2$ to $x_1$. The maximum value $x_1$ can take is $0.2624$ due to the constraint (\ref{eqn_new_lp3}). Thus, for $\mathbf{x}^\mathrm{T}=[0.2624 \textup{ } 0.1458 \textup{ } 0.5039 \textup{ } 0.0879]^\mathrm{T}$, the highest $z$ in the region becomes $\mathbf{b}^{\mathrm{T}}\mathbf{x}=52.4265$. Moreover, 
\begin{align}
z &= \mathbf{b}^{\mathrm{T}}\mathbf{x}= 67x_1+17(0.4082-x_1) + 59\times0.5039  \nonumber \\
&+ 30\times0.0879 \implies x_1=\frac{z-39.3065}{50}, \nonumber \\
x_2&=0.4082-\frac{z-39.3065}{50} = \frac{59.7165-z}{50}, \nonumber \\
x_3&=0.5039, \textup{ } x_4=0.0879.
\end{align}

\item At the end of the previous region, we have $\mathbf{x}^\mathrm{T}=[0.2624 \textup{ } 0.1458 \textup{ } 0.5039 \textup{ } 0.0879]^\mathrm{T}$. Notice that this is the exact corner point of the open polytope of the feasible set (\ref{eqn_new_lp1})-(\ref{eqn_new_lp5}) that gives the optimal solution to the first problem. Thus, for the region $z>52.4265$, the new constraint $\mathbf{b}^{\mathrm{T}}\mathbf{x} \leq z$ is nonbinding, which means the solution is exactly the same as that of the first problem.
\end{itemize}

\begin{table}
\begin{center}
\caption{Solution of the Third Optimization Problem}
\label{tab_opt3_first}
\begin{tabular}{|c|c|c|c|c|c|}
\hline
& $x_1$ & $x_2$ & $x_3$ & $x_4$ & Capacity \\ \hline
$z=35$ & $0$ & $0.4082$ & $0.3554$ & $0.2364$ & $0.8288$ \\ \hline
$z=45$ & $0.1139$ & $0.2943$ & $0.5039$ & $0.0879$ & $0.8469$ \\ \hline
\end{tabular}
\end{center}
\end{table}

The results of solving the third optimization problem at specific $z$ values (average adder size values) are given in~Table~\ref{tab_opt3_first}.

\section{Online Learning} \label{online}

In the online learning approach, during the operation of the device, error rates are collected and the device controller decides the optimal switching points. Here, the controller uses the collected data to find the BER functions via curve fitting again. However, since it is online learning, training data is different as it is not collected over the entire device lifespan of interest. Recall that the solution of the first problem in offline learning provides the switching points optimal for average storage capacity. Here in online learning, curve fitting will be used to obtain different BER functions, but we still consider solving a problem similar to the first offline problem to obtain the optimal switching points during the operation of the device. We experimented with four different practical setups. These setups differ in how the training data is collected.

In the below setups, we are aided by the offline learning results. We assume here that such results are obtained from a device that differs from the one being used.

\subsection{First Setup}
In this setup, we benefit from the switching points found by offline learning in Section~\ref{offline} to determine a way of collecting data points online. The device collects data at three different regions, where the first (resp., second and third) region has a width of $5\%$ of the TDMR density range of interest such that the end point of this region is the first (resp., second and third) switching point determined by offline learning. In every region, the device collects six equally distant BER data points in the respective region and uses this data to obtain a BER function with curve fitting. Then, by equating this function to the BER threshold, it finds the optimal switching point.

\subsection{Second Setup}
In this setup, we benefit from the switching points found by offline learning in Section~\ref{offline} to determine a way of collecting data points. The device collects data at three different regions, where the first (resp., second and third) region has a width of $5\%$ of the TDMR density range of interest such that the end point of this region is $5\%$ of the TDMR density range away from the first (resp., second and third) switching point determined by offline learning. In every region, the device collects six equally distant BER data points in the respective region and uses this data to obtain a BER function with curve fitting. Then, by equating this function to the BER threshold, it finds the optimal switching point.

\subsection{Third Setup}
In this setup, we again benefit from the switching points found by offline learning in Section~\ref{offline} to determine a way of collecting data points. The device collects data at three different regions, where the first (resp., second and third) region has a width of $10\%$ of the TDMR density range such that the end point of this region is $5\%$ of the TDMR density range away from the first (resp., second and third) switching point determined by offline learning. In every region, the device collects six equally distant BER data points in the respective region and uses this data to obtain a BER function with curve fitting. Then, by equating this function to the BER threshold, it finds the optimal switching point.

\subsection{Fourth Setup}
Here, we use the switching points determined by offline learning to create four regions in the TDMR density range. These regions, called main regions, occupy $25\% - 15\% - 45\% - 15\%$ of the TDMR density range, $[0.8,1.5]$. These percentages are determined based on the offline learning switching points. Then, within every main region, we specify a training region such that this region has a width of $10\%$ of the TDMR density range and its end point is determined randomly with a condition that its end point should be at most $20\%$ of the TDMR density range away from the main region end point. After determining the training regions, we collect six equally distant BER data points in the respective region and use this data to obtain a BER function with curve fitting. Then, by equating this function to the BER threshold, we find the optimal switching point (done by the device controller). Table~\ref{fitting_coef_on} shows the coefficients of the polynomials resulting from curve fitting for this setup.

\begin{table}
\begin{center}
\caption{Coefficients of the Fifth-Degree Polynomials Resulting From Curve Fitting for Fourth Online Learning Setup}
\label{fitting_coef_on}
\begin{tabular}{|l|l|l|l|}
\hline
\multicolumn{1}{|c|}{} & \multicolumn{1}{c|}{\textbf{$m_{i,1}$}} & \multicolumn{1}{c|}{\textbf{$m_{i,2}$}} & \multicolumn{1}{c|}{\textbf{$m_{i,3}$}} \\ \hline
\textbf{OP-LOCO}       & $-0.276$                            & $1.463$                             & $-2.905$                                         \\ \hline
\textbf{SP-LOCO}       & $0.150$                            & $-0.866$                             & $1.992$                                              \\ \hline
\textbf{OT-LOCO}       & $0.019$                            & $-0.090$                             & $0.172$                                                     \\ \hline
                       & \multicolumn{1}{c|}{\textbf{$m_{i,4}$}} & \multicolumn{1}{c|}{\textbf{$m_{i,5}$}}  & \multicolumn{1}{c|}{\textbf{$m_{i,6}$}}         \\ \hline
\textbf{OP-LOCO}       & $2.783$                             & $-1.304$                              & $0.241$                                         \\ \hline
\textbf{SP-LOCO}       & $-2.256$                            & $1.252$                               & $-0.272$                                              \\ \hline
\textbf{OT-LOCO}       & $-0.167$                            & $0.082$                             & $-0.016$                                                   \\ \hline
\end{tabular}
\end{center}
\end{table}

\subsection{Numerical Results}
Table~\ref{tab_online_res} shows the results of the online learning setups. We can observe that the random setup (the fourth setup) gives slightly inferior average storage capacity compared with the other three setups. The main reason for this capacity difference is that the other three setups use the knowledge obtained from offline learning differently via the switching points. The second setup differs from the first setup as a result of the separation from the offline switching points. The third setup differs from the second setup because the width of the training regions is larger, which affects curve fitting.

\begin{table}
\begin{center}
\caption{Results of Online Learning}
\label{tab_online_res}
\begin{tabular}{|c|c|c|c|c|c|}
\hline
Setup & $x_1$ & $x_2$ & $x_3$ & $x_4$ & Capacity \\ \hline
First setup & $0.2451$ & $0.1816$ & $0.4762$ & $0.0971$ & $0.8538$ \\ \hline
Second setup & $0.2626$ & $0.1842$ & $0.4871$ & $0.0661$ & $0.8583$ \\ \hline
Third setup & $0.2547$ & $0.1847$ & $0.4822$ & $0.0784$ & $0.8565$ \\ \hline
Fourth setup & $0.2443$ & $0.1799$ & $0.4684$ & $0.1074$ & $0.8527$ \\ \hline
\end{tabular}
\end{center}
\end{table}

\begin{remark}
We can observe, aided by Fig.~\ref{fig_perf1}, that the results of online learning lead to exceeding the BER middle track threshold of $10^{-3}$. This is due to the fact that online learning setups do not have all the fitting points over the device lifetime to obtain an optimal solution that prohibits constraint violation for all TD density values. However, this violation occurs in just $1.85\%$ of the TD density range $[0.8,1.5]$ for the first setup, $6.06\%$ for the second setup, $4.07\%$ for the third setup, and just $1.60\%$ for the fourth setup. Depending on the application, this violation can be tolerable.
\end{remark}

\section{Simulation results}\label{sim_res}

In this section, we discuss the gains achieved via optimal reconfiguration and compare different learning setups. We also compare our results to an idea from recent literature \cite{ahh_otloco} and to a straightforward idea. Data points of the ST-LOCO coding scheme are shown in the figures for completeness.

\begin{figure}
\center
\includegraphics[trim={0.5in 0.7in 0.6in 0.9in}, width=3.3in]{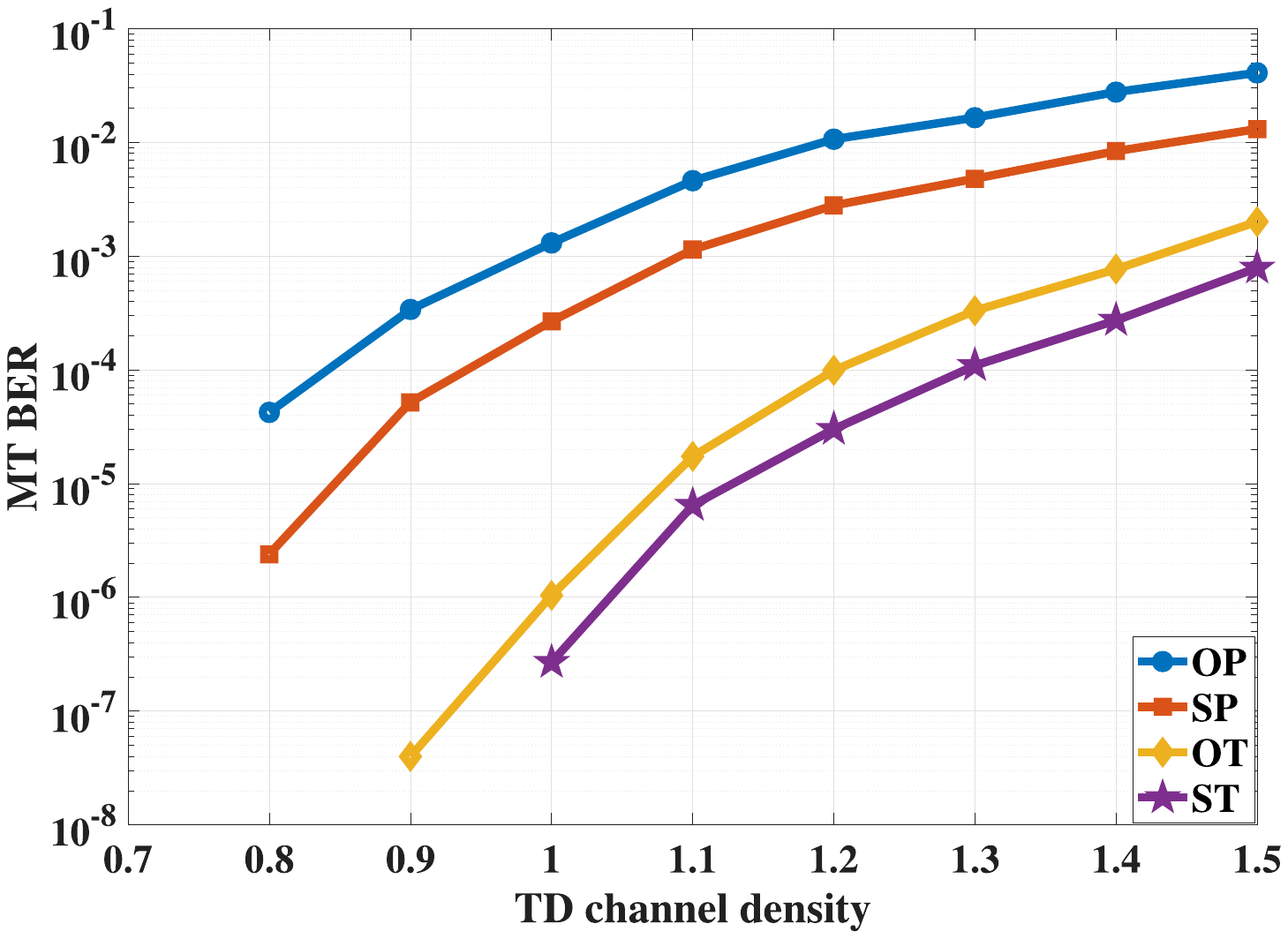}
\caption{Middle-track BER performance of all coding schemes.}
\label{fig_perf1}
\end{figure}

\begin{figure}
\center
\includegraphics[trim={0.5in 0.3in 0.6in 0.9in}, width=3.3in]{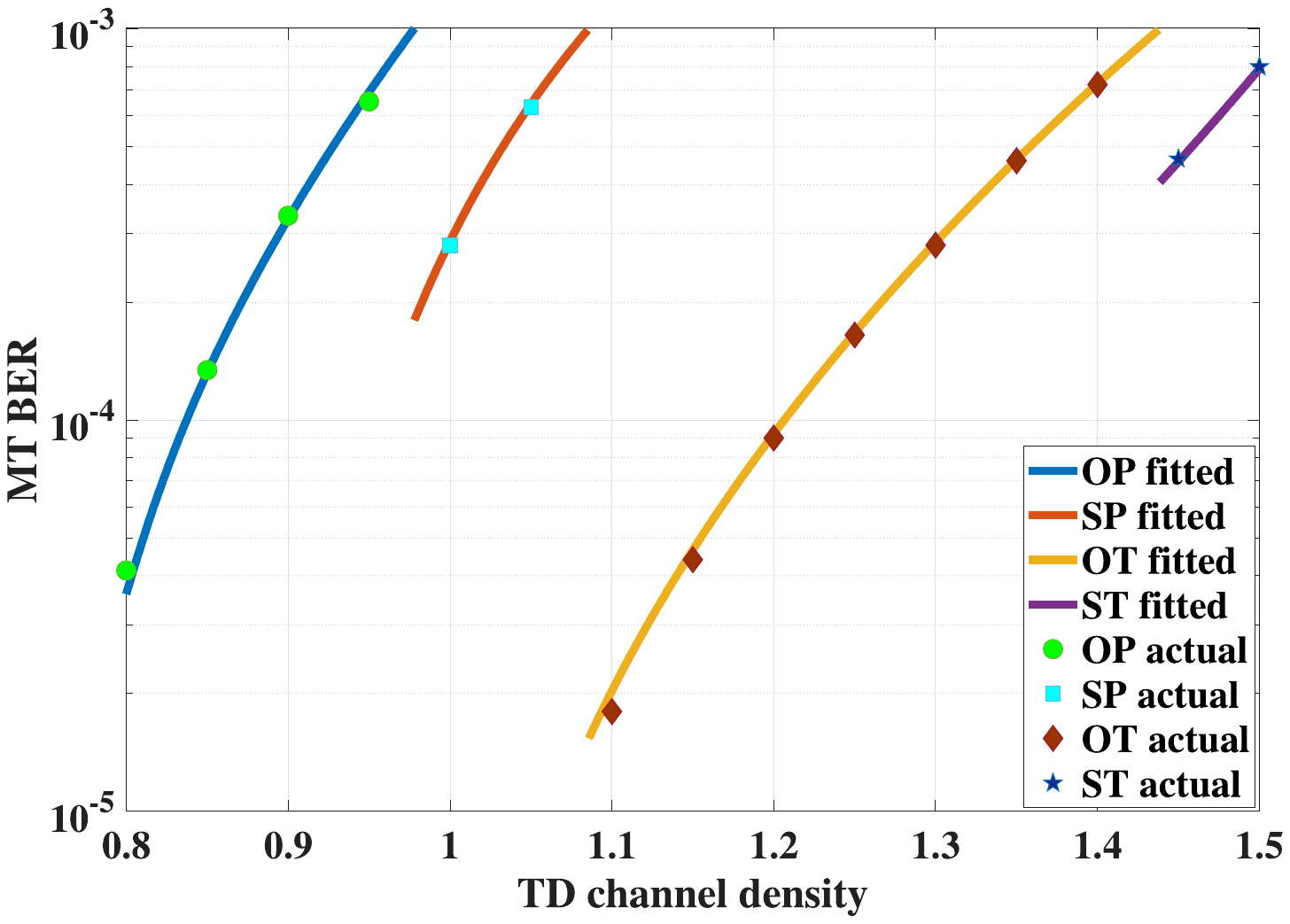}
\vspace{-1.2em}
\caption{BER performance of the first offline learning setup.}
\label{fig_perf2}
\vspace{-0.2em}
\end{figure}

\begin{figure}
\center
\includegraphics[trim={0.5in 0.7in 0.6in 0.9in}, width=3.3in]{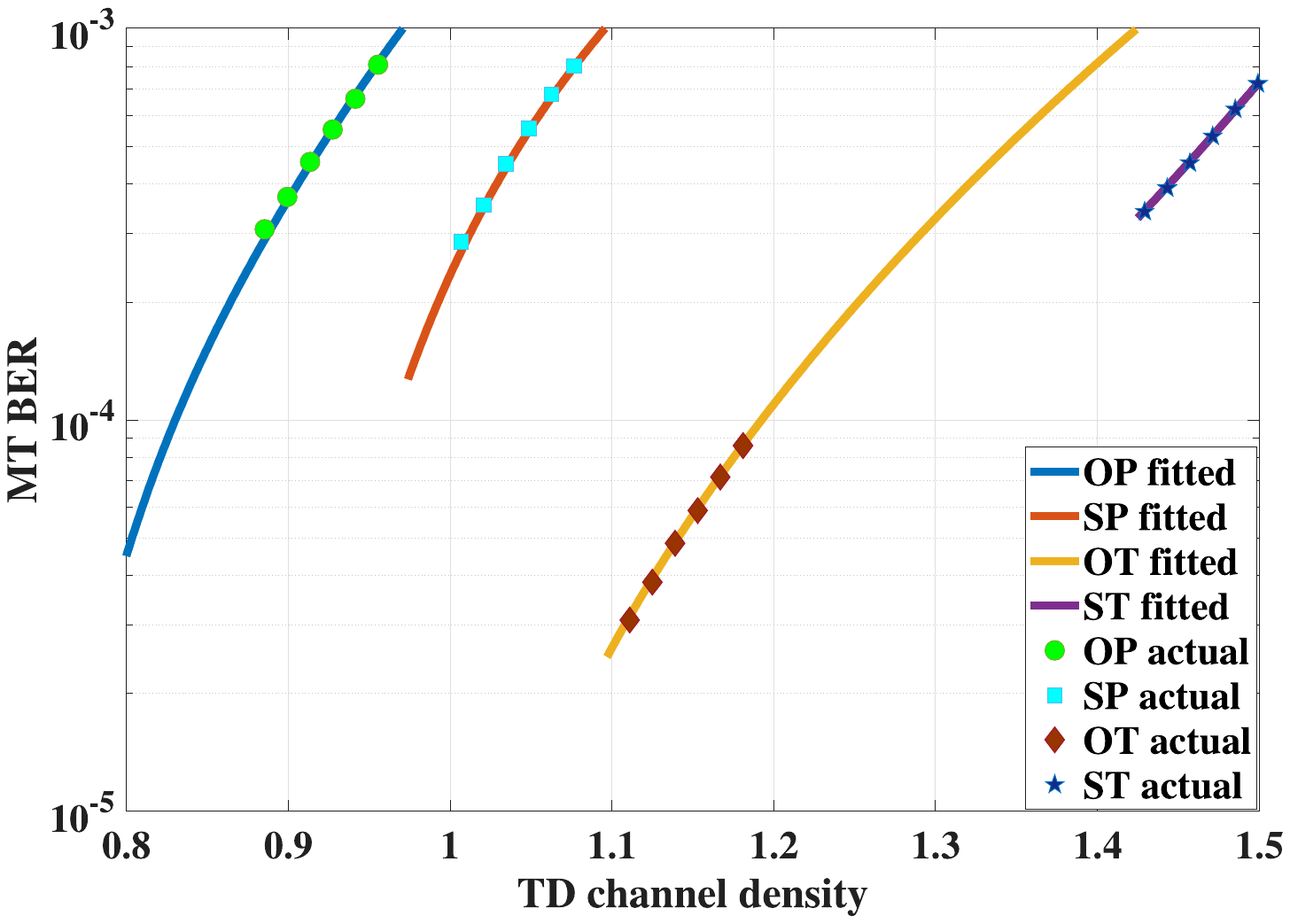}
\caption{BER performance of the fourth online learning setup.}
\label{fig_perf3}
\end{figure}

First, we compare the offline learning method and the online learning method. We choose the first offline learning setup and the fourth online learning setup as our comparison candidates. Fig.~\ref{fig_perf2} and Fig.~\ref{fig_perf3} show the middle-track (MT) BER performance of offline learning and online learning, respectively. Fig.~\ref{fig_perf2} shows that each coding scheme except ST-LOCO uses all the available density region until its performance reaches the BER threshold. On the other hand, in Fig.~\ref{fig_perf3}, online learning offers acceptable BER performance with a rate loss of $0.20\%$. As a result of the incomplete data online learning suffers from, it may switch between coding schemes at a point earlier than needed, which results in storage capacity loss compared with offline learning, or at a point later than needed, which results in BER constraint violation. Having said that, even though online learning does not have all the training data, it can provide promising storage capacity and BER performance with the guidance of the offline learning. We also note that the available ST-LOCO training data indicates that ST-LOCO BER reaches the threshold of $10^{-3}$ at a TD density of $1.528$. Depending on the application needs, this might mean that storage device will not be able to offer reliable performance after this point.

After comparing offline and online learning, it is also worth noting that offline learning exhausts one block of the storage device to collect the data throughout the lifetime range of interest, but it offers more accurate decisions assuming all blocks behave the same over time. Online learning, on the contrary, does not require exhausting a device block, and it better responds to differences between blocks.

Now, we compare our offline learning result of the first setup with an idea that decides the switching points manually, based on predetermined time/density stamps. In our previous work, we obtained the switching points manually such that the middle-track BER is roughly kept under $10^{-3}$ \cite{ahh_otloco}, without any accurate fitting or consideration of the device status. The idea in our previous work \cite{ahh_otloco} gives an average adder size of $61.9$ and a capacity of $0.8638$ as it uses OP-LOCO and OT-LOCO coding schemes. Our proposed offline learning setup with BER shown in Fig.~\ref{fig_perf2} gives an average adder size of {$52.43$ and incurs a capacity loss of $1.09\%$. These results tell that with a limited capacity loss, one can reduce the complexity remarkably, $15.3\%$ reduction here, via smaller average adder size using offline learning. Moreover, our setup with BER shown in Fig.~\ref{fig_perf2} offers a better BER performance and a longer lifetime. This is due to the fact that the idea in our previous work is not able to maintain the middle-track BER under $10^{-3}$ around the switching points or at the end of the density range, unlike the offline learning setup. This means if the comparison metrics are the same within a fixed lifetime, our offline learning setup would offer a better storage capacity at a smaller adder size compared with this idea. Note that the third offline learning setup can even offer further complexity gains by bounding the average adder size.

Next, in our proposed optimization problem, we set the BER threshold to the highest error rate observed in the idea of \cite{ahh_otloco} to enable a fair comparison. This highest error rate is $1.6 \times 10^{-3}$, and it is the BER of the OT-LOCO coding scheme at TD density $1.5$ as shown in Fig~\ref{fig_perf1}. With this configuration, we obtained the results shown in Table~\ref{tab_new_thres} for the first offline learning setup. Observe that under a fair comparison, our new optimization-based approach increases the capacity and decreases the complexity via smaller average adder size compared with the idea in \cite{ahh_otloco}. Another interpretation of this result is that if the device lifetime is kept the same, our approach offers better BER performance. Furthermore, if the usage BER threshold is kept the same at $1.6 \times 10^{-3}$, our approach with optimal reconfiguration increases the device lifetime by nearly $10.67\%$ (from TD density $1.5$ to $1.66$).

\begin{figure}
\center
\includegraphics[trim={0.5in 0.7in 0.6in 0.9in}, width=3.3in]{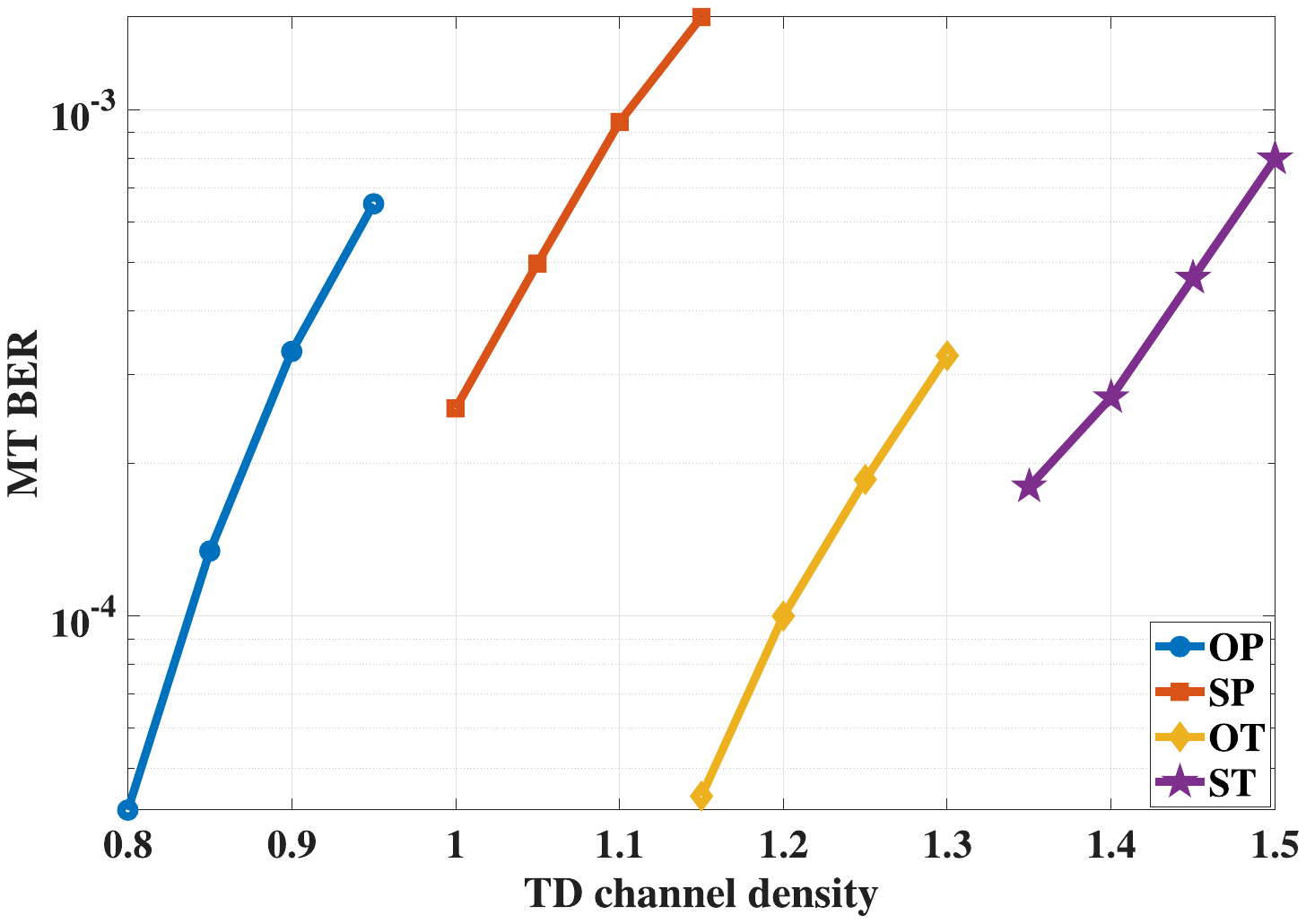}
\caption{BER performance of the idea that every coding scheme has the same share of the device lifetime.}
\label{fig_perf4}
\end{figure}

\begin{table}
\vspace{-0.2em}
\begin{center}
\caption{Shares of Coding Schemes, Capacity, and Adder Size in First Setup of Offline Learning With Threshold $1.6 \times 10^{-3}$}
\label{tab_new_thres}
\begin{tabular}{|c|c|c|c|c|c|}
\hline
$x_1$ & $x_2$ & $x_3$ & $x_4$ & Capacity & Adder size \\ \hline
$0.3036$ & $0.1711$ & $0.5253$ & $0$ & $0.8674$ & $54.24$ \\ \hline
\end{tabular}
\end{center}
\end{table}

Finally, we compare our new results based on offline and online learning with the straightforward idea of dividing the lifetime equally among coding schemes. Fig.~\ref{fig_perf4} shows the middle-track BER performance when each coding scheme is utilized $25\%$ of the lifetime. Observe that this setup violates the middle-track BER constraint, since the BER surpasses $10^{-3}$, in around $7\%$ of the TD density range $[0.8, 1.5]$, while our offline learning fully satisfies the constraint and our online learning, fourth setup, just violates it in $1.60\%$ of the range. Here, in the straightforward idea, the average storage capacity is $0.8452$ and the average adder size is $43.25$. Consider the third offline learning setup with $z=43.25$. This setup achieves an average storage capacity of $0.8451$. In particular, our third offline learning setup achieves the same average storage capacity and adder size as those of the straightforward idea while fixing the problem of constraint violation, which demonstrates the effectiveness of our approach. If the BER threshold is raised to match the highest BER reached by the straightforward idea, our third offline learning setup will achieve notable reduction in average redundancy ($4\%$--$5\%$).

\begin{remark}
The results of the second optimization problem demonstrate the complexity reduction our approach can offer. In Table \ref{tab_opt1}, the setup with $x_1=x_3=0$ offers an average adder size of $24.69$. When compared with the average adder size offered in \cite{ahh_otloco},  $61.9$, and the average adder size offered by the straightforward idea of dividing lifetime equally, $43.25$, we observe the remarkable gains achieved by the optimization approach that aims to minimize the complexity.
\end{remark}

\begin{remark}
As a result of adopting a wide read head, the middle track receives additional protection via LOCO coding. Ideally, the BER should be consistent across all TDMR down tracks, including the middle one. However, due to unequal data protection, this balance is disrupted, resulting in the difference between all-tracks and middle-track BERs. Having said that, all-tracks performance using our approach is still promising.
\end{remark}

\begin{remark}
Even though we developed the concept of reconfiguration to optimize storage capacity and complexity for LOCO coding schemes in TDMR devices with wide read heads, our methods and techniques are broadly applicable. They can be extended to different coding schemes, such as LDPC codes, and adapted for various applications, including Flash memory and wireless communication systems.
\end{remark}

\section{Conclusion}\label{sec_conc}

We introduced a novel learning-based approach for reconfiguring the LOCO coding scheme optimally in TDMR systems. In particular, we designed three optimization problems for offline learning that aim to maximize storage capacity, maximize storage capacity as well as minimize complexity, and maximize storage capacity while keeping complexity bounded, respectively. We showed that these optimization problems are convex and solved them via KKT conditions and linear programming fundamentals. After deriving the optimal solutions, we presented the associated practical insights. We also designed online learning procedures for reconfiguration, which aim to maximize storage capacity but they determine the switching points while the TDMR device is operating. We numerically showed the notable gains achieved via offline learning and online learning methods compared with a recent-literature idea and a straightforward one. We showed the BER performance plots of the proposed methods. We noted that offline learning offers remarkable gains due to the fact that it is designed according to the training data collected over the whole lifetime interval. Online learning, however, offers promising results even though it does not have access to all training data. Future work includes combining our LOCO codes with multi-dimensional LDPC codes \cite{ahh_md} for TDMR systems and using machine learning to reconfigure between modern rate-compatible LDPC codes \cite{ahh_rmc_sc}.

\section*{Acknowledgment}

The authors would like to thank Mohsen Bahrami, Prof. Bane Vasi\'{c}, and Beyza Dabak for providing and developing the TDMR system model that we updated and used to generate the results in Section~\ref{offline}, Section~\ref{online}, and Section~\ref{sim_res}. The authors would also like to thank Duru Uyar for her assistance in carrying out this research.
\ifCLASSOPTIONcaptionsoff
\newpage
\fi
\end{document}